\documentclass[aps,twocolumn,superscriptaddress,showpacs]{revtex4}
\usepackage{amssymb}
\usepackage{graphicx}

\input{tcilatex}

\begin{document}

\title{Metal-to-semiconductor transition in squashed armchair carbon
nanotubes}
\author{Jun-Qiang Lu}
\affiliation{Center for Advanced Study, Tsinghua University, Beijing 100084, P. R. China}
\affiliation{Department of Physics, Tsinghua University, Beijing 100084, P. R. China}
\author{Jian Wu}
\affiliation{Center for Advanced Study, Tsinghua University, Beijing 100084, P. R. China}
\author{Wenhui Duan}
\affiliation{Department of Physics, Tsinghua University, Beijing 100084, P. R. China}
\author{Feng Liu}
\affiliation{Department of Materials Science and Engineering, University of Utah, Salt
Lake City, Utah 84112}
\author{Bang-Fen Zhu}
\affiliation{Center for Advanced Study, Tsinghua University, Beijing 100084, P. R. China}
\author{Bing-Lin Gu}
\affiliation{Center for Advanced Study, Tsinghua University, Beijing 100084, P. R. China}
\affiliation{Department of Physics, Tsinghua University, Beijing 100084, P. R. China}
\date{\today}

\begin{abstract}
We investigate electronic transport properties of the squashed armchair
carbon nanotubes, using tight-binding molecular dynamics and Green's
function method. We demonstrate a metal-to-semiconductor transistion while
squashing the nanotubes and a general mechanism for such transistion. It is
the distinction of the two sublattices in the nanotube that opens an energy
gap near the Fermi energy. We show that the transition has to be achieved by
a combined effect of breaking of mirror symmetry and bond formation between
the flattened faces in the squashed nanotubes.
\end{abstract}

\maketitle



The discovery of carbon nanotubes\cite{SI} has stimulated intensive research
interests, partly because of their unique electronic properties and their
potential application in nanodevices. In particular, much effort has been
made to manipulate the low-energy electronic properties of carbon nanotubes,
as it is the requisite step for using nanotubes to realize a functional
device.

A single-walled nanotube (SWNT) can be either semiconducting or metallic
depending sensitively on its diameter and helicity.\cite{White}
Specifically, a tube is metallic if the Fermi $K$-point of the corresponding
graphene sheet, from which the tube is wrapped, remains to be an allowed $k$%
-point by the periodic boundary condition for the tube; otherwise, it is
semiconducting. Consequently, most previous studies have focused on a
popular idea of modifying the electronic properties of SWNTs by structural
perturbation, in an attempt to shift the Fermi $k$-point away from an
allowed state, resulting in a metal-to-semiconductor transition (MST).

Various experimental methods, such as twisting\cite{KM}, introducing
topological defects\cite{CCB}, and stretching\cite{CCR}, have been used to
manipulate the electronic and transport properties of nanotubes. Theoretical
studies\cite{CJP,AM,PEL} have also been performed to help exploring the
correlation between the structural perturbation and the change of electronic
properties. However, in general, the experiments are done in a guess-work
manner, because it is \textit{a prior} unknown how a given structural
perturbation would change the electronic properties. One major difficulty is
that the structural perturbation occurs for atoms in real-space, but the
change of electronic properties has to be revealed by distribution of
k-points of electronic bands in reciprocal-space.

\begin{figure}[b]
\begin{center}
\includegraphics[width=3.0in]{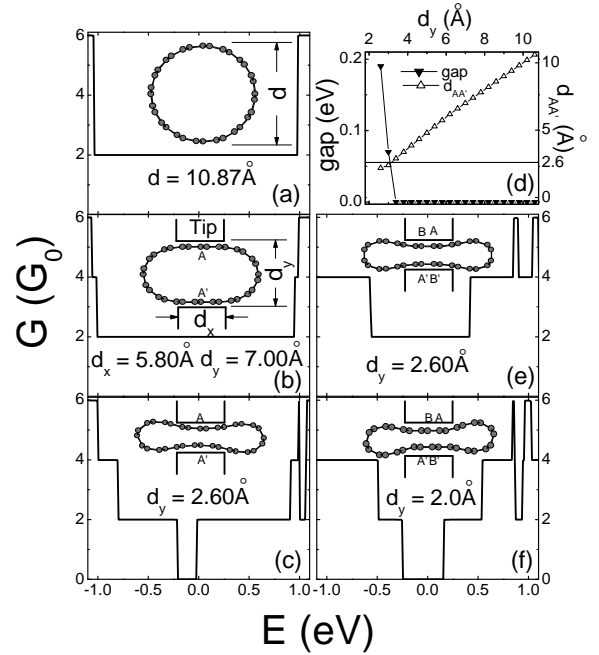}
\end{center}
\caption{ (a)-(c),(e),(f) Conductances of various nanotube structures, which
are shown as the insets. $E$ is the energy of injected electrons, and the
Fermi energy of ideal armchair (8,8) nanotube is taken as zero. (d) The
conductance gap and $d_{AA^{\prime }}$ as a function of the tip distance $%
d_{y}$.}
\end{figure}

In this Letter, we demonstrate a new method of manipulating electronic
properties of SWNTs by examing directly the atomic structural perturbation
in real space without the need of analyzing the distribution of $k$-points
in reciprocal space. We show that when a structural perturbation makes the
two original equivalent sublattices in the tube distinguishable, it will
open an energy gap for a metallic armchair SWNT, leading to a MST. This can
be achieved, for example, by simply squashing the nanotubes. Furthermore, we
show that the \textit{physical} distinction of the two sublattices must be
achieved by a combined effect of mirror symmetry breaking (MSB) and bond
formation between the flattened faces of the squashed tubes, while neither
the MSB nor the bonding alone would result in the MST.

We demonstrate the basic principles of our method by squashing an armchair
(8,8) SWNT, as shown in Fig. 1. The simulations, for both structural
optimization and calculation of electronic transport property, are performed
using four-orbital tight-binding (TB) method. To squash the tube, two
identical tips with a width of $d_{x}=5.80$\AA\ are used to press the tube
symmetrically about its center in the $\pm y$ direction, as shown in Fig.
1b. The tips are assumed to be super stiff with a hard-wall interaction with
the tube. At each tip position, the atomic structure, i.e., the shape of the
tube is optimized with a TB molecular dynamics\cite{CHX} code. Most
noticeably, as the two tips moves towards each other, the tube cross section
changes from a circle to an ellipse (Fig. 1b), and then to a dumbbell (Fig.
1c). As long as the distance between the two tips is not too short ($>1.8$%
\AA ), the tube is found to maintain its structural integrity, and the whole
process is reversible. Further pressing the tips to shorter distances ($<1.8$%
\AA ) would permanently damage the tube.

Using the optimized structure at all tip positions, we employ the TB Green's
function method\cite{JCC,JW,JW2} to study the electronic transport
properties of the squashed tubes, which are modeled as a typical
left-lead--conductor--right-lead system. Within the framework of Landauer
approach, the conductance is expressed as 
$G=G_{0}\text{Tr}(\Gamma _{L}\mathcal{G}_{C}\Gamma _{R}\mathcal{G}%
_{C}^{\dag})$, 
\cite{HC,SD} where $G_{0}(=2e^{2}/h)$ is the unit quanta of conductance, $%
\mathcal{G}_{C}$ is the Green's function of the conductor, and $\Gamma _{L}$
and $\Gamma _{R}$ are the spectral density describing the coupling between
the leads and the conductor through the imaginary component of self-energy.

The typical conductance curve of a perfect armchair (8,8) SWNT is shown in
Fig. 1a. It represents a metallic behavior, which is well-known for armchair
SWNTs. The conductance near the Fermi energy is $2G_{0}$, indicating that
there are two conducting channels. For the squashed tube, we consider two
different cases: one breaking the mirror symmetry (MS) about the $y$-axis
(Figs. 1b and 1c) and the other preserving the MS (Fig. 1e).

When the tube is squashed without MS, its conductance remains at $2G_{0}$
near the Fermi energy with an elliptical shape (Fig. 1b), but drops sharply
to zero with a dumbbell shape (Fig. 1c). Thus, a MST can be achieved by
squashing the tube, but only after the tube is squashed to a dumbbell shape.
The physical difference is that the two flattened faces of an elliptical
tube remain separate without bonding (atomic-orbital overlap); while they
become close enough in a dumbbell tube to form new bonds (see discussion
below). It has been suggested that the MSB may lead to opening an energy gap
in a metallic armchair SWNT\cite{CJP}. However, Fig. 1b clearly shows that
the MSB by itself can not open up an energy gap and its only effect is to
cause a slight variation in the conductance step. A gap may only be opened
after the atomic orbitals on the two flattened faces of the squashed tube,
without MS, overlap with each other to form new bonds.

\begin{figure}[b]
\begin{center}
\includegraphics[width=1.8in]{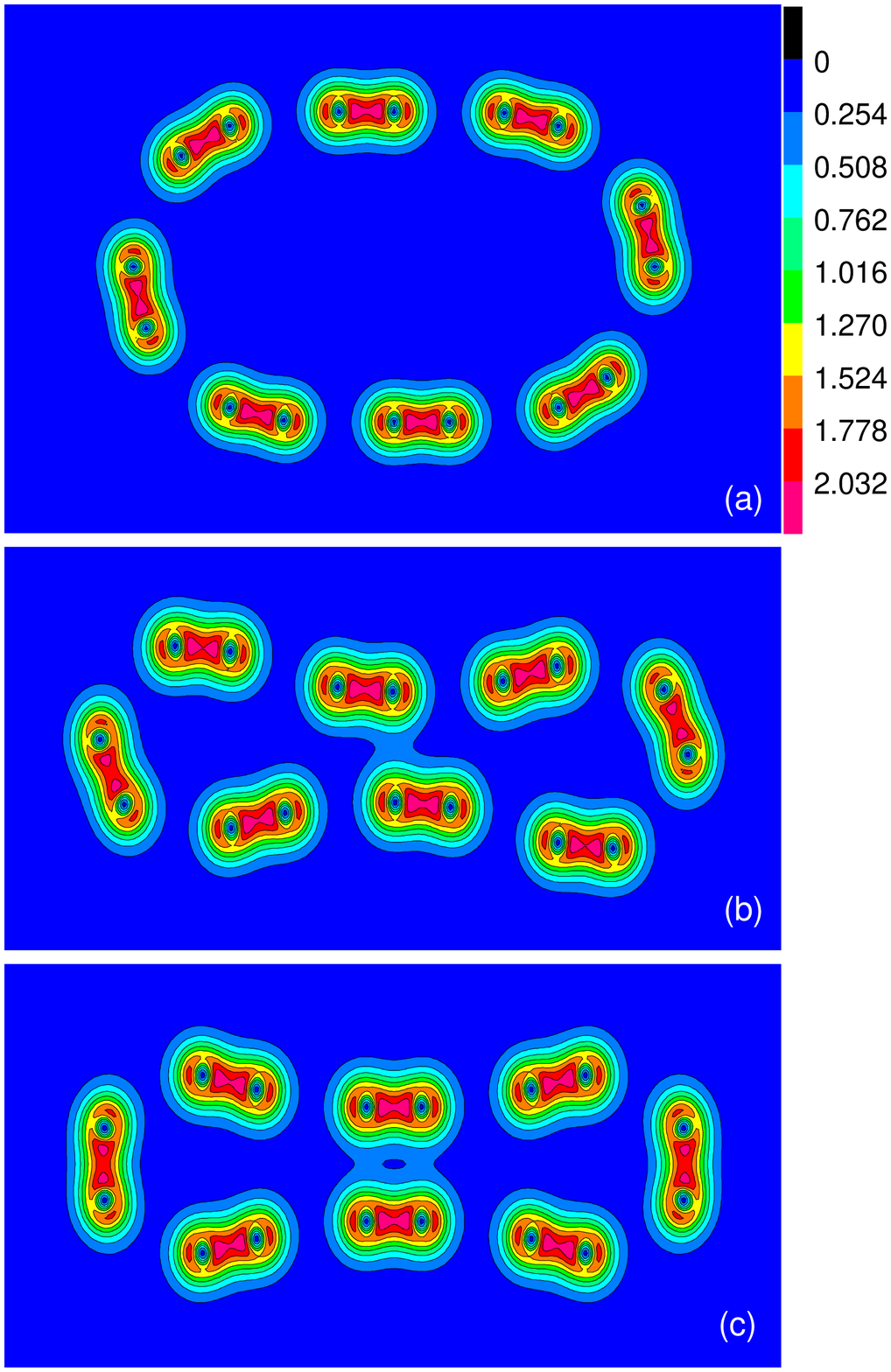}
\end{center}
\caption{(color) Contour plots of the charge density (unit: $e/$\AA $^{3}$)
in the cross section of the nanotube structures in Figs. 1b, 1c, and 1e.}
\end{figure}

To quantify the degree of squashing in terms of the bonding (atomic-orbital
overlap) between the two flattened faces, we monitor the distance between
the two closest atoms, $d_{AA^{\prime }}$, A in the upper face and A' in the
lower face, as shown in Fig. 1b. In Fig. 1d, we plot the conductance gap
near the Fermi energy and the interaction distance $d_{AA^{\prime }}$, as a
function of the tip separation $d_{y}$. 
Clearly, the conductance gap appears when $d_{AA^{\prime }}<2.6$\AA .
Because a cutoff length of $2.6$\AA\ for the C-C bond is used\cite{JCC} in
the TB calculation, it indicates that the gap is only opened after the atom
A starts to form bond with the atom A'. This is further confirmed by
plotting of charge density, as shown in Fig. 2. It can be vividly seen that
the charge density overlaps between the two flattened faces in the dumbbell
tube (Fig. 2b), reflecting the new bonding between the atoms A and A'. In
contrast, no density overlap, and hence bonding in the elliptical tube (Fig.
2a).

The above results of the squashed tubes without MS demonstrate that the bond
formation between the flattened faces plays an important role in driving the
MST. However, it remains unknown whether such bonding alone is sufficient to
induce the MST, i.e., whether the MSB also plays a role, as suggested before.%
\cite{CJP,PD} To test this, we take a look at the squashed tubes preserving
the MS, as shown in Fig. 1e. Interestingly, the conductance remains at $%
2G_{0}$ near the Fermi energy, even when the distance between the two
flattened faces is less than $2.60$\AA . This indicates that the MST can not
be induced by the bonding, if the MS is preserved. (The bonding between the
two flattened faces is also reflected by the distribution of charge
densities in Fig. 2c.) Thus, we can conclude that the MST can be driven by
neither the MSB nor the bonding alone; it has to be driven by the combined
effect of the two.

Next, we show that the combined effect of the MSB and the bonding between
the flattened faces in a squashed armchair SWNT is to make the two original
equivalent sublattices in the tube distinguishable, and such distinction can
then be used as a unique condition for driving the MST. It is well-known
that the graphene sheet and hence the nanotube have two equivalent
sublattices, which we may label as A and B sublattices. The operation of
squashing can then be defined in reference to the two sublattices. If the
y-axis, along which we squash the tube, is chosen to pass through two atoms
from the same A-sublattice (or B-sublattice), as the case in Figs. 1b and
1c, the squashed tube will break the MS about the y-axis. Upon the atomic
orbital overlap between the two flattened faces, A-atoms bond with A-atoms
(Fig. 2b). If the y-axis is chosen to pass through two atoms from different
sublattices (one from A and the other from B), as the case in Fig. 1e, the
squashed tube will maintain the MS. Upon the atomic orbital overlap,
A(B)-atoms bond with B(A)-atoms (Fig. 2c). In the following, we will refer
to the first case as the AA' structure (Figs. 1c and 2b) and the second as
the AB' structure (Figs. 1e and 2c). Note the two differs by a rotation of
about $7.5^{{\circ }}$ of their respective y-axes.

\begin{figure}[t]
\begin{center}
\includegraphics[width=2.80in]{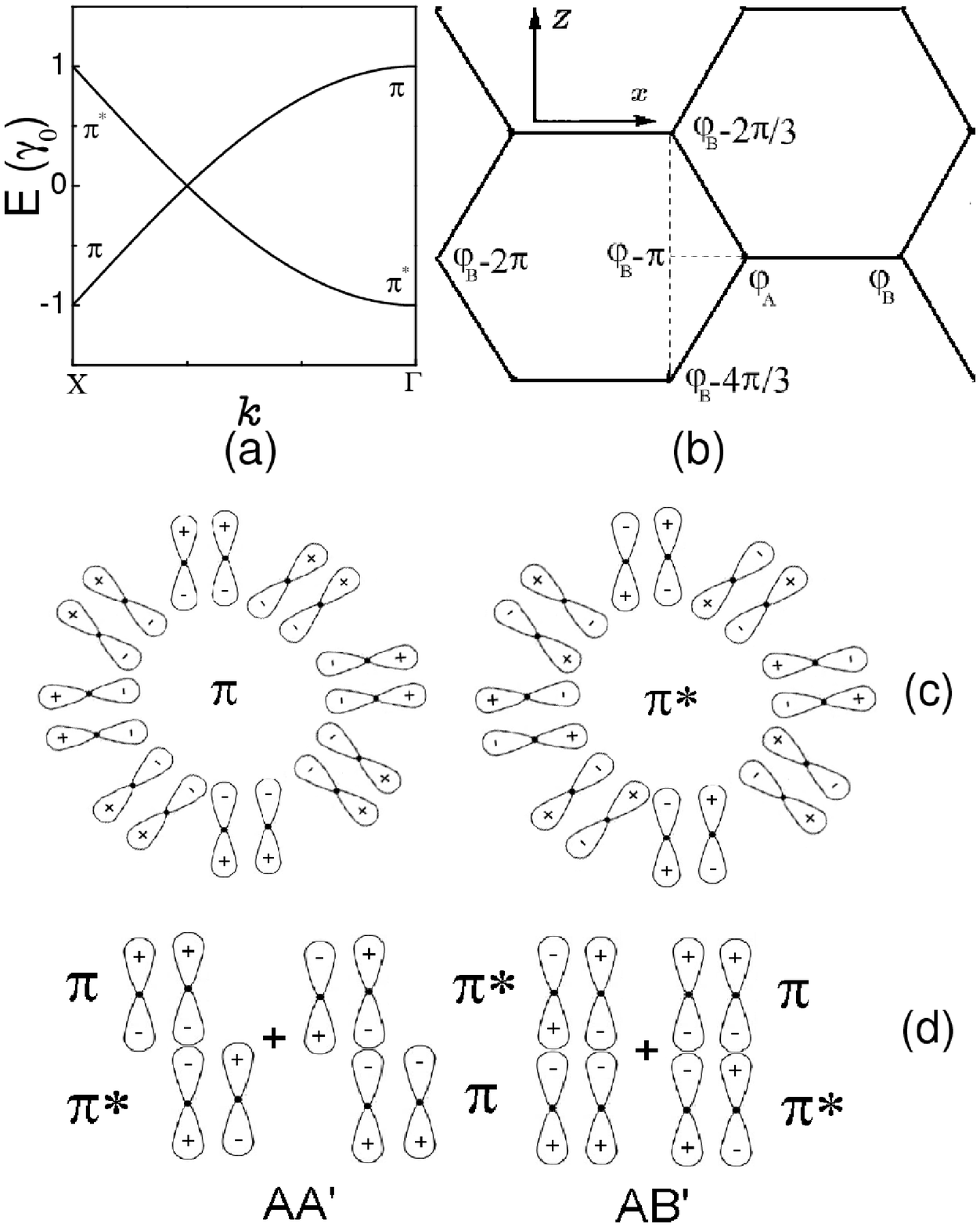}
\end{center}
\caption{ (a) The energy dispersion relations near the Fermi energy of an
ideal armchair (8,8) SWNT with a $pp\protect\pi $ model. (b) The phase
correlations at the Fermi vector $k_{F}$ between the three equivalent atomic
positions B which are the nearest neighborhoods of the atomic positions A.
(c) A schematic representation of the states $\protect\pi $ and $\protect\pi %
^{\ast }$ within the cross section of an ideal archair (8,8) SWNT with $q=8$%
. (d) Configurations of the new bonds formed by atomic orbital overlap
between the two states $\protect\pi $ and $\protect\pi ^{\ast }$ for
structures AA' and AB'. The AA' structure consists of two $\protect\sigma $
bonds between A and A'; the AB' structure consists of two $\protect\sigma $
bonds and two $\protect\sigma $ antibonds betwee AB' and A'B. }
\end{figure}

For an ideal armchair (8,8) SWNT, its metallic behavior can be understood
from its energy dispersion relations within a simple $pp\pi $ model\cite{RS}%
: 
\[
E_{q}(k)={\pm }{\gamma _{0}}\left\{ 1 \pm 4\cos (q\pi /8)\cos (ka/2)
+4\cos^{2}(ka/2)\right\} ^{1/2} , 
\]
\begin{equation}
(-\pi <ka<\pi ; q=1,...,8),
\end{equation}
where $k$ is the wave vector along the $z$ axis, 
$a=2.46$\AA\ is the lattice constant, and $\gamma _{0}$ is the nearest
neighborhood hopping integral. The energy dispersion relations near the
Fermi energy are shown in Fig. 3a. The two lines, crossing at the Fermi
point, correspond to the two eigenstates (bonding $\pi $ and antibonding $%
\pi ^{\ast }$) with the quantum number $q=8$, as all 8 atoms in one
sublattice have the same phase. The phase relations between the nearest
neighborhood atoms at the Fermi vector $k_{F}(=\pm 2\pi /3)$ are shown in
Fig. 3b. The states $\pi $ and $\pi ^{\ast }$ within the tube cross-section
is schematically shown in Fig. 3c. Note that the interaction energies
between the two sublattices cancel out each other by symmetry, leading to a
zero total interaction energy: $\gamma_{0}e^{i(\varphi_{B}-\varphi
_{A})}(1+e^{-i2\pi /3}+e^{-i4\pi /3})=0$. This cancellation, independent of
the phase difference $(\varphi _{B}-\varphi _{A})$, leads to the degeneracy
of the eigenstates $\pi $ and $\pi ^{\ast }$ at $k_{F}$.

We next extend the above model by including the interaction (bonding)
between the two flattened faces in a squashed nanotube, using a perturbation
Hamiltonian\cite{PD} 
\begin{equation}
H^{\prime }(k)=\left( 
\begin{array}{cc}
\delta _{\pi \pi }(k) & \delta _{\pi \pi ^{\ast }}(k) \\ 
\delta _{\pi ^{\ast }\pi }(k) & \delta _{\pi ^{\ast }\pi ^{\ast }}(k)%
\end{array}%
\right) .
\end{equation}
The diagonal matrix elements $\delta _{\pi \pi }$ and $\delta _{\pi
^{\ast}\pi ^{\ast }}$ merely act to shift the location of the $\pi$ and $\pi
^{\ast}$ bands, and hence the energy and location of band crossing. It is
the off-diagonal elements $\delta _{\pi \pi ^{\ast }}$ and $\delta _{\pi
^{\ast }\pi }$ that cause quantum mechanical level repulsion and hence open
up an energy gap.

If a MS exists, like in the AB' structure, the mirror operator $M$ must be
applied, we have $M(\pi )=\pi$; $M(H^{\prime})=H^{\prime}$; $M(\pi
^{\ast})=-\pi ^{\ast}$. Then, $\delta _{\pi \pi ^{\ast }}=M(\delta _{\pi \pi
^{\ast}})=M(\langle \pi |H^{\prime }|\pi ^{\ast }\rangle )=-\langle \pi
|H^{\prime}|\pi ^{\ast}\rangle =-\delta _{\pi \pi ^{\ast }}$, which gives $%
\delta _{\pi \pi^{\ast}}=0$. Thus, if the MS is preserved, the off-diagonal
elements are always zero and the band crossing persists without gap opening,
regardless of whether a bond exists between the two flattened faces. This
indicates that the MSB is a necessary (but not sufficient) condition for the
MST.

A schematic representation of the bonding configuration between the two
states $\pi$ and $\pi ^{\ast }$ for the two structures AA' and AB' is shown
in Fig. 3d. For the AA' structure, the off-diagonal element $\delta _{\pi
\pi^{\ast }}$ consists of two $\sigma $ bonds as $\langle
p|H^{\prime}|p\rangle +\langle p|H^{\prime}|p\rangle \neq 0$, where $%
|p\rangle $ is the $2p_y$ orbital of carbon atom. In contrast, for the AB'
structure, the off-diagonal element consists of four $\sigma $ bonds, which
cancel out as $\delta _{\pi \pi ^{\ast}}=(-\langle p|H^{\prime
}|p\rangle+\langle p|H^{\prime }|p\rangle)+(\langle p|H^{\prime }|p\rangle
-\langle p|H^{\prime }|p\rangle )=0$. So, the off-diagonal element for the
AB' structure is always zero, in agreement with the mirror-symmetry analysis
discussed above.

\begin{figure}[t]
\begin{center}
\includegraphics[width=3.0in]{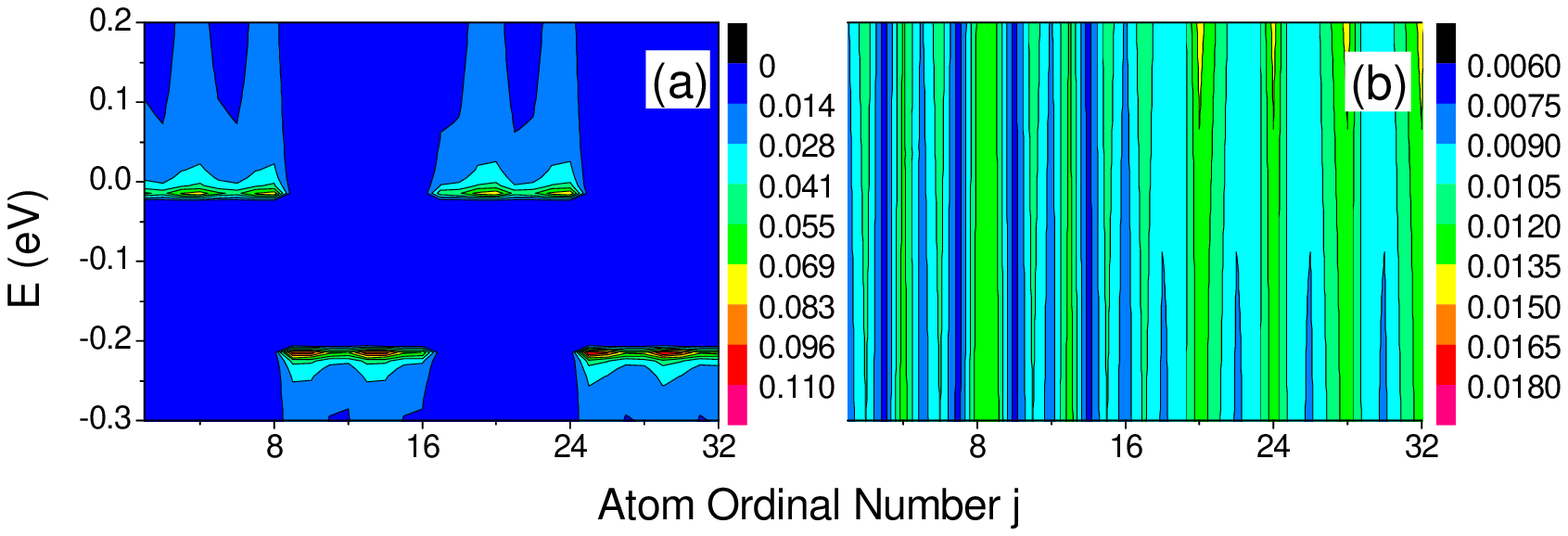}
\end{center}
\caption{(color) The LDOS (unit: eV$^{-1}$) distributions near the Fermi
energy over two atomic layers of the nanotubes for AA' (a) and AB' (b)
structures. The atoms in B (A) sublattice are labled 1 (9) through 8 (16)
for the first atomic layer and 17 (25) through 24 (32) for the second atomic
layer. }
\end{figure}

The above analysis clearly demonstrates that the MST must be driven by a
combination of the MSB and the bond formation, which effectively
distinguishes the two originally equivalent sublattices (A and B). Without
the bonding, the two sublattices are always equivalent. Upon the bonding,
the two remains equivalent if the MS is preserved because the bonding occurs
between atoms from two sublattices in a symmetric manner, but becomes
distinguishable if the MS is broken because the bonding occurs between atoms
within only one sublattice (A), making it different from the other
sublattice (B). Such a distinction of the two sublattices is further
revealed by the local density of states (LDOS), as shown in Fig. 4. The LDOS
is defined by the Green's function as\cite{JW} 
$\text{LDOS}(j,E)=-\frac{1}{\pi }{\text{Im}}\mathcal{G}_{C}(j,j,E)$, 
where $j$ is the atom index in the nanotube. In Fig. 4, the LDOS
distributions of two atomic layers (along z-axis) are plotted at an energy
interval from $-0.3$eV to $0.2$eV. Each layer consists of two sublattices (A
and B) of 16 atoms. For an ideal tube, the LDOS near the Fermi energy are
homogeneously distributed over the two equivalent sublattices A and B. For a
squashed tube, in a AA' structure (Fig. 4a), the bonding between atoms A and
A' distinguishes the A sublattice from the B sublattice, resulting in a
redistribution of the LDOS. Specifically, the electrons tend to distribute
around A sites below the Fermi energy, but around B sites above it. This
causes a discontinuity in the energy spectrum, as shown in Fig. 4a. In
contrast, in AB' structure (Fig. 4b), the LDOS crosses the Fermi energy
continuously because the off-diagonal elements are zero and the states $\pi $
and $\pi ^{\ast }$ continue to be the eigenstates. The inhomogeneity of the
LDOS in the Fig. 4b is caused by the inhomogeneous curvature of the squashed
nanotube.

Last, we show that squashing the armchair SWNT can be used as a general
approach to drive the MST, which is practically important. It would be
rather inconvenient, if the MST can only be driven by squashing the tube
along a specific direction breaking the MS. Fortunately, we find that the
MST can in fact be driven by squashing the tube along any direction. In case
one starts with squashing the tube along a direction that preserves the MS,
all it needs to be done is to squash the tube to a larger extent, to where a
spontaneous symmetry breaking occurs. Fig. 1f shows that for AB' structure,
if one continues to press the tips beyond Fig. 1e ($d_{y}=2.60$\AA ) to Fig.
1f ($d_{y}=2.00$\AA ), a gap near the Fermi energy will eventually appear,
because of the spontaneous breaking of MS, causing the two sublattices
distinguishable.

In summary, we demonstrate that squashing the armchair SWNT can be used as a
general approach to induce a MST, which may find practical applications in
novel nanodevices, such as used for a mechanical nano-switch. The underlying
mechanism is to make the two originally equivalent sublattices in the tube
distinguishable, which requires a combined effect of MSB and bond formation
between the two flattened faces in the squashed tube. Such distinction of
two sublattices is likely to be generally responsible for the semiconductor
behavior of certain classes of nanotubes, such as BN nanotubes. Besides
squashing the tube, other methods, such as chemical adsorption, might be
used to distinguish the two sublattices and hence to induce the MST.

The work is supported by the Ministry of Education of China, the National
High Technology Research and Development Program of China (Grant No.
2002AA311153). Liu thanks support from US-DOE (Grant No. DE-FG03-01ER45875)
and the Natural Science Foundation of China (Grant No. 69928403). 

\end{document}